%Paper: hep-th/9308069
%From: Samir Mathur <me@ctpdown.mit.edu>
%Date: Sat, 14 Aug 93 22:34:29 EDT

%%%%%%%%%%%%%%%%%%%%%%%%%%%%%%%%%%%%
%There is one figure appended to the paper after the end which
%should be separated and printed. This requires (on a unix machine)
%`uudecode', then `uncompress', then printing the postscript file.
%%%%%%%%%%%%%%%%%%%%%%%%%%%%%%%%%%%%
%%%%%%%%%%%%%%%%%%%%%%%%%%%%%%%%%%%%%%%%%%%%%%%%%%%%%%%%%%%%%%%%%%%%% 
%%%%
\input harvmac
\overfullrule=0 pt
\def\x{x^+}
\def\xn{x^+_0}
\def\xnp{{x^+_0}'}

\Title{}
{{Information Retrieval from a Charge `Trap'}}
\centerline{{\bf Samir D. Mathur}}

\bigskip\centerline{Center for Theoretical Physics}
\centerline{Massachussets Institute of Technology}
\centerline{Cambridge, MA 02139}
\vskip .7cm

We study the model of massless $1+1$ electrodynamics with nonconstant  
coupling,
introduced by Peet, Susskind and Thorlacius as the `charge hole'. But  
we take
the boundary of the strong coupling region to be first timelike, then  
spacelike
for  a distance $X$, and then timelike again (to mimic the structure  
of a black
hole).  For an incident charge pulse entering this `charge trap' the  
charge and
information get separated. The charge comes out near the endpoint of  
the
singularity. The `information' travels  a well localised path through  
the
strong coupling region and comes out later.

\vskip .1 in
\Date{\hfill  August 1993}

\newsec{Introduction.}

The recent renewal of interest in the black hole  information paradox  
has
brought in new ideas, as well as a revival of some old ones. In  
\ref\CGHS{C.
Callan, S. Giddings, J. Harvey and A. Strominger, Phys. Rev. {\bf D  
45} (1992)
1005.} a simple $1+1$ dimensional model of the black hole evaporation  
process
was formulated. The gravity sector consists of the metric and a  
dilaton, and in
$1+1$ dimensions has no propagating degrees of freedom. The matter is  
taken to
be massless scalar fields.  These can be  integrated out
explicitly to leave an effective action for the dilaton-gravity  
system that
incorporates quantum backreaction on the geometry. One can then  
pursue
semiclassical analysis of this effective action.

One result from this line of investigation has been the realisation  
that
information can be effectively `lost' by being trapped for long times
deep in the black hole. The large number of remnants that one can
create this way do not violate unitarity of loop amplitudes or normal
thermodynamic experience because the the degrees of freedom
in the deep `throat' of the black hole do not interact appreciably  
with
observers far outside the hole \ref\BD{T. Banks, A. Dabholkar, M.R.  
Douglas and
M. O'Loughlin, Phys. Rev. {\bf D 45} (1992) 3607.}. For the case of  
holes with
charge,
it is possible to transport information to regions of spacetime from  
which no
classical paths lead to the asymptotic regions of the original  
spacetime,
all without encountering any region of strong gravitatonal coupling  
\ref\TRI{S.
Trivedi, Phys. Rev. {\bf D 47} (1993) 4233,  A. Strominger and S.  
Trivedi,
Preprint NSF-ITP-93-15 (hep-th 9302080). }.

An opposing viewpoint is one seeking complete information recovery in
the Hawking evaporation process. Peet, Susskind and Thorlacius  
\ref\PST{A.
Peet, L. Susskind and L. Thorlacius, Phys. Rev. {\bf D 46} (1992)  
3435.}
studied $1+1$ electrodynamics in a linear dilaton background. The  
dilaton
causes the electromagnetic coupling constant to vary as   
$g^2=e^{-2x}$. A left
moving fermion charge pulse, on entering the strong
coupling region, gets reflected to a right moving pulse. The pulse  
suffers
distortion (there is created  a trail of electron positron pairs) but  
all the
information sent in at ${\cal I}^-_R$ (past null infinity on the  
right) is
recovered at ${\cal I}^+_R$ (future null infinity on the right).   
Thus the
incident charge pulse forms a temporary `charge hole' which then  
evaporates.
We will for convenience use the term `charge hole' for this time  
dependent
solution, though the authors of \PST\ had used the term for the case  
where the
electromagnetic field was maintained as  a fixed background rather  
than  as a
dynamical field.This model
 illustrates how information is recovered when the full  
nonperturbative matter
- gauge field  interaction is considered, even
though a perturbative treatment of the interaction leads to  
information
loss into the strong coupling region.

In \ref\RST{J.G. Russo, L. Susskind and L. Thorlacius, Phys. Rev {\bf  
D 46}
(1992) 3444, Phys. Rev. {\bf D 47} (1993) 533.} it was proposed that  
for the
dilaton gravity plus matter system, all fields (matter, gravity,  
dilaton)
reflect off the strong dilaton region. This boundary condition  
offered the best
hope of maintaining some form of cosmic censorship in the model. It  
was noted
however that imposing reflection makes sense only when the strong  
dilaton
boundary is timelike. Independent assumptions are needed to obtain  
physics  in
the causal future of the `singularity' at which the strong dilaton  
boundary is
spacelike.

In \ref\VV{E. Verlinde and H. Verlinde, Preprint PUPT-1380,  
IASSNS-HEP-93/8, K.
Schoutens, E. Verlinde and H. Verlinde, Preprint PUPT-1395,  
IASSNS-HEP-93/25.}
the boundary condition of \RST\ was extended to a full quantum theory  
of
dilaton gravity plus matter. Reflection in the strong coupling region  
gives the
complete physics of black hole formation and evaporation.
The boundary condition is applied along a line that is timelike in a  
fixed
background coordinate region. This line extends
into the strong coupling region, going to the coordinate region  
behind the
semiclassical location of the singularity.
 It is argued that
strong fluctuations of spacetime in the neighbourhood of the  
singularity
invalidate the question of whether the singularity itself is timelike  
or
spacelike (a semiclassical analysis would say that it is spacelike).

Unfortunately even with such a reflection condition all information  
does not
emerge  at  ${\cal I}^+_R$. The Hilbert space is a direct sum of two  
parts, and
one part disappears into the black hole. One might seek to modify the  
boundary
condition to obtain full information retrieval. But it would be most
satisfactory if no boundary condition was needed at all; rather the  
reflection
was actually produced by the nature of the physics in the strong  
coupling
region.  The charge hole of \PST\ , for instance, had reflected  
incident waves
without an explicit boundary condition.

In this paper we return  to the charge hole (i.e. dilaton  
electrdynamics) and
ask the following question. We know that a spatially varying coupling
$g^2=e^{-2x}$ prevents penetration  of information into the strong  
coupling
region ($x\rightarrow -\infty$). What happens if the wavepacket  
enters a region
of spacetime where the coupling is increasing with time?  This  
corresponds to a
spacelike strong coupling boundary, like the singularity of  a black  
hole.  In
this situation the state cannot simply avoid the domain of strong  
coupling by
reflecting to a weak coupling region.
We wish to see the behavior of some measure of `information' when the
wavepacket encounters such a spacelike boundary. Even though the  
strong
coupling physics could be very different  between the charge and the  
gravity
cases, we hope by this study to get some feeling for how
`information' might move at strong coupling, and what kind of states  
are
created when matter is forced through such regions.

To address the above question we modify the charge hole of \PST\ to a   
`charge
trap' described as follows. Let $(t,x)$ be the usual Minkowski  
coordinates on
the $1+1$ spacetime, and let $x^\pm=t\pm x$.
(Our metric signature is $(1,-1)$.)   We let
\eqn\ONEONE{\eqalign{g^2~=&~e^{-2x} ~~~{\rm for} ~~x^+\le 0\cr
=&~e^{2t} ~~~{\rm for} ~~0\le x^+\le X\cr
=&~e^{-2(x-X)} ~~~{\rm for} ~~X\le x^+\cr}}
$g^2$ is continuous throughout spacetime. But the boundary of strong  
coupling,
defined by $g^2(t,x)=1$, is first timelike ($x=0$ for $-\infty<t<0$),
then spacelike ($t=0$ for $0<x<X$) and then timelike again ($x=X$
for  $0<t<\infty$). This shape of the boundary mimics the strong  
coupling
boundary of a black hole.  The first timelike section corresponds to  
the time
before black hole formation, the spacelike section is the  
singularity, and the
last timelike section is the trajectory of the end point of black  
hole
evaporation, possibly a naked singularity.  The term `charge trap'  
represents
the fact that in this model some wavepackets can be forced to  
encounter a
region where the coupling increases in  timelike direction, so that  
they cannot
simply reflect back to the weak coupling region in the way they did  
in the
charge hole.

The line where $g^2=1$ is sketched in fig. 1.  We expect that the  
wavepacket
marked `A',  moving left with $x^+<<-X$, would reflect off the $x=0$  
line in
the manner described in \PST\  and return to the weak coupling  
region.
 Similarily wavepacket `B', with $x^+>>X$ would reflect off the line  
$x=X$.
 But wavepacket `C', with $-X<x^+<0$ would reflect off the line $x=0$  
and enter
a domain where the coupling becomes strong in a timelike direction.   
Wavepacket
`D', with $0<x^+<X$, would directly enter the region where the  
coupling became
strong in a timelike direction.  What is the future evolution of  
wavepackets
`C' and `D'?

 Even with $g^2$ given by \ONEONE\ the model is exactly solvable.
 We define one measure of `information' and follow this for an  
incident charge
pulse during its evolution   through
 various regions of spacetime.  For the behavior of the information,   
a priori
there are  three possibilities:

 (i)\quad If the wavepacket is `expelled' from the strong coupling  
region
 even when the boundary is spacelike, then the information in  
wavepackets `C'
and `D' should travel out along the singularity and escape to    
${\cal I}^+_R$
from the endpoint of the singularity. The escaping pulse would be the  
analogue
of the `thunderbolt' in the black hole case. Such propagation along  
the
singularity will be acausal, and has been called a `conveyer belt  
transport'
along the singularity.

 (ii)\quad The wavepacket can enter the strong coupling region behind  
the
singularity and continue deep into it, never coming out. This would  
give
 an eternal remnant.

 (iii)\quad The wavepacket may initially be forced to enter the  
strong coupling
region, but may ultimately escape to  ${\cal I}^+_R$, sometime after  
the strong
coupling line has reverted back to being timelike (i.e., after
 the evaporation of the hole). This would be information leaking out   
over time
from the  remnant or naked singularity.

 In section 2 of this paper we review the charge hole model. In  
section 3 we
analyse the flow of information in different wavepackets in the  
`charge trap'
 described above. Section 4 is a discussion.

\newsec{The charge hole.}

The Schwinger model, which describes a massless fermion coupled to  
the
electromagnetic field in $1+1$ dimensions, is remarkable in that it  
can be
solved exactly \ref\SCH{J. Schwinger, Phys Rev. {\bf 128} (1962)  
2425.}. The
field content turns out to be a single massive scalar field, with the  
mass
dependent on the electromagnetic coupling constant. Peet et al.  
consider this
model coupled to a background dilaton field $\phi$. The action is
\eqn\TWOONE{I~=~\int d^2x[i\bar \psi
\gamma^\mu(\partial_\mu+iA_\mu)\psi-{1\over
4}e^{-2\phi(x)}F_{\mu\nu}F^{\mu\nu}]}
Thus the dilaton field gives a nonuniform electromagnetic coupling
$g^2(x)=e^{2\phi(x)}$. The Schwinger model continues to be exactly  
solvable,
even with nonuniform coupling. The fermi field can be bosonised:
\eqn\TWOTWO{\eqalign{\bar\psi\gamma^\mu\psi~=~&j^\mu~~\rightarrow~~
{1\over \sqrt{\pi}}\epsilon^{\mu\nu}\partial_\nu Z\cr
\psi_L~~\rightarrow~~ &:e^{i\sqrt{4\pi} Z_L}:\cr
\psi_R~~\rightarrow~~ &:e^{i\sqrt{4\pi} Z_R}:\cr}}
with $Z_L$, $Z_R$ the left and right moving parts of the boson
\eqn\TWOTHREE{Z_{L,R}~=~{1\over 2}[Z~\mp~\int_x^\infty dx' \partial_0  
Z(x')]}

In terms of $Z$ the action \ONEONE\ becomes
\eqn\TWOFOUR{I~=~\int d^2x[-{1\over 2} \partial_\mu Z\partial^\mu  
Z~-~
{1\over  \sqrt{4\pi}}\epsilon^{\mu\nu}F_{\mu\nu}Z~-~{1\over 4 g^2(x)}
F^{\mu\nu}F_{\mu\nu}]}
Integrating out the vector potential gives
\eqn\TWOFIVE{I~=~\int d^2x[-{1\over 2} \partial_\mu Z\partial^\mu  
Z~-~
{g^2(x)\over 2\pi} Z^2]}

An initial state of definite energy on  ${\cal I}^-_R$ is given as
\eqn\TWOSIX{|{\rm in}>~=~\int dx^+e^{-ip_+x^+ }\psi_L(x^+)|0>~=~
\int dx_0^+e^{-ip_+x_0^+}:e^{i\sqrt{4\pi}Z_L(x_0^+)}:|0>}
Here $|0>$ is the `in' vacuum.  Thus we need to study coherent states
$:e^{i\sqrt{4\pi}Z_L(x_0^+)}:|0>$ of the free boson field, which  
means  that we
need to study the classical evolution of the field $Z$. The classical
configuration corresponding to the above coherent state is a left  
moving wave
at  ${\cal I}^-_R$:
\eqn\TWOSIXP{Z_C=\sqrt{\pi}\theta(x^+-x_0^+)}
 Let $g^2=e^{-2x}$. The classical equation of motion is
\eqn\TWOSEVEN{\partial_+\partial_-Z_C~=~-{1\over 4\pi}g^2  
Z_C~=~-{1\over
4\pi}e^{x^-}e^{-x^+}Z_C}
Let $e^{-x^+}=u$, $e^{x^-}=v$. This gives
\eqn\TWOEIGHT{\partial_u\partial_v Z_C~-~{1\over 4\pi}Z_C~=~0}
Thus we obtain the free wave equation with constant (tachyonic) mass  
term in
these coordinates.  For the step function profile at ${\cal I}^-_R$
the full classical solution is
\eqn\TWONINE{Z_C~=~\sqrt{\pi}\theta(x^+-x_0^+)J_0[{1\over \sqrt{\pi}}
e^{x^-/2}(e^{-x_0^+}-e^{-x^+})^{1/2}]}
We get the asymptotic out state on ${\cal I}^+_R$ by taking the limit
$x^+\rightarrow \infty$ in \TWONINE .
\eqn\TWOEIGHTP{Z_C~\rightarrow~\sqrt{\pi} J_0[{1\over \sqrt{\pi}}  
e^{x^--\xn}]}
 We find that no part of the state reaches ${\cal I}^+_L$. By  
contrast,
classically the left moving pulse would have continued to move  
leftwards, and
would reach ${\cal I}^+_L$.
In the full quantum theory the chiral anomaly couples the left and  
right moving
fermion modes, so that a left moving wavepacket gets converted to a  
right
moving one on reaching the strong coupling region.
Perturbation theory in the coupling would have revealed the anomaly,  
but some
part of the left moving wave would still reach ${\cal I}^+_L$. The  
function
\TWOEIGHTP\ is interpreted \PST\ as describing a right moving charge  
pulse with
a tail of fermion - antifermion pairs.

One can now recover the scattering amplitude  for arbitrary boson  
states, which
can then be rewritten to get the scattering matrix for fermion  
states. We will
not concern ourselves much with these steps, which are outlined in  
\PST\ and go
through in an identical manner for our modification of the model.

Note that the single fermion state \TWOSIX\ corresponds to  
superposing the
$\theta$-functions \TWOSIXP\ with different amplitudes for different  
$\xn$, but
this is not the same as adding together the classical  
$\theta$-functions
with the  different coefficients. A fermion - antifermion pair  
corresponds to a
 superposition of `bump' functions
$\sqrt{\pi}\theta(\x-{\x}_1)-\sqrt{\pi}\theta(\x-{\x}_2$).  On the  
other hand
an arbitrary classical function $Z_C$ corresponds to a complicated  
nonlocal
construct in the fermion language. We shall ignore the issues  
involved with
interpreting the states arising from the refermionisation of $Z$, and  
study
information flow for the boson field only. Since $Z$ is a free field,  
we need
study only its classical solutions.

We drop the subscript $C$ on $Z_C$ in what follows, since we shall be  
studying
these classical solutions at all times.

\newsec{Information flow in the charge trap.}

Our goal is to keep track of the `information' flow as classical  
solutions
$Z(t,x)$ evolve in the charge trap defined in section 1.  How do we  
define
`information'? One conserved quantity that is exhibited by the  
solution $Z$ is
the fermion charge density $j^0={1\over \sqrt{\pi}}{\partial Z\over  
\partial
x}$. For the initial condition \TWOSIXP\  the charge density at very  
early
times  is a delta function at  $x^+=x^+_0$; the total charge is  
unity.  The
charge moves left to the strong coupling region, and is then expelled  
back to
large $x$.  One might try to track the information in the solution by
computing the classical `charge outside distance $x$':
\eqn\THREEONE{Q(x)~\equiv~\int_x^\infty dx'j^0(x')~=~\int_x^\infty  
dx'
{1\over \sqrt{\pi}} {\partial Z\over \partial x'}~=~{1\over
\sqrt{\pi}}(Z(\infty)-Z(x))}

But, as we shall see, this charge does not  give a good measure of  
the flow of
information in the solution $Z$.  We consider instead the following  
way of
defining information.  For a free scalar field, we have one other
conserved quantity besides the above charge: the Wronskian of two  
solutions of
the field equation
\eqn\THREETWO{(Z_2,Z_1)~\equiv~\int dx~[ Z_2^*(x)\partial_0Z_1(x)~-~
\partial_0Z_2^*(x) Z_1(x)]}
\eqn\THREETHREE{\partial_0(Z_2,Z_1)~=~0}
If we find the Wronskian between two functions to be nonzero, then we  
can
discriminate between the functions.   Thus we will define  
`information flow' by
tracking the region that contributes   to the Wronskian  of a pair of  
solutions
$Z_1$, $Z_2$ of the free wave equation.  More specifically, we will  
consider a
pair of solutions which at   ${\cal I}^-_R$ have the form
\TWOSIXP\ appropriate to localised single fermion initial states:
\eqn\THREEFOUR{\eqalign{Z_1~\rightarrow~\sqrt{\pi}\theta(x^+-x_0^+)\c 
r
Z_2~\rightarrow~\sqrt{\pi}\theta(x^+-{x_0^+}')\cr}}
The Wronskian
\eqn\THREEFOURP{(Z_2,Z_1)~=~\pi~\int d\x  
\{\theta(\x-\xnp)\delta(\x-\xn)
{}~-~\delta(\x-\xnp)\theta(\x-\xn)\}}
 is $-\pi$ if $\xnp>\xn$ and $\pi$ if  $\xnp<\xn$.
The information in $Z_1$ is localised at $\xn$ in the following  
sense:  If we
keep only the part of the function in a small neighbourhood of  
$\x=\xn$ then we
can deduce correctly the Wronskian with any  function built from
$\theta$-functions $Z_2$ with $\xnp<\xn$.  We cannot however say  
anything
about the Wronskian with $\theta$-functions having $\xnp>\xn$.

On examining the waveforms evolving from \THREEFOUR\   we will find  
the
following. There is always  a region  for each function $Z_i$, with  
width order
unity, such that  restricting the function to this region is just  
like
restricting the $\theta$-functions to a neighbourhood of their `jump'   
(for the
 purpose of computating  the Wronskians). That is to say, by keeping  
only the
form of $Z_1$ in the relevant region we can obtain correctly the  
Wronskian with
all functions $Z_2$ which had $\xnp<\xn$, while  we get just zero if
 $\xnp>\xn$. It is interesting that the `feature' of the functions  
$Z_i$
travels in a well localised region, even though this feature moves  
far  away
from its initial trajectory $\x={\xn}_i$. Thus by studying Wronskians  
we track
the progress of this `feature' through different regions of  
spacetime, and call
it the information in the initial state.

For our following claculations, we let $\xnp>\xn$, with $\xnp-\xn$  
not much
smaller than unity. Here the scale `unity' is set by the rate of  
increase of
the coupling, which had the form $e^{-2x}$ in the charge hole, and  
has a
similar rate of change in the charge trap \ONEONE .  Thus we do not  
focus on
the behavior of very short distance scales, i.e. high momenta,  in  
the initial
state.  We expect that such high momentum components would penetrate  
deeper
into the strong coupling region. Our principal interest is in the  
effect of the
spacelike singularity of length $X$, and we wish to separate the  
effect of high
momenta and of  large $X$ in studying the penetration of states into  
the strong
coupling region.

The significance of locating the `information' in this manner is the  
following.
If there is a region of spacetime in which the information does not  
enter, then
no significant effect should be produced if we alter the physics in  
that
region. (We may delete that region altogether and place a reflection  
condition
at the created boundary.) But if we wish to delete  a region through  
which
information does flow, then the  boundary conditions have to  
reproduce to some
approximation the flow of information found  in the complete  
spacetime.

\subsec{The charge hole ($g^2=e^{-2x}$).}

Let us  first see how information moves in the charge hole studied in  
\PST ,
i.e. for $g^2=e^{-2x}$.  The solutions for initial conditions  
\THREEFOUR\ are
\eqn\THREESIX{\eqalign{Z_1~=~&\sqrt{\pi} \theta(\x-\xn)J_0[{1\over  
\sqrt{\pi}}
e^{x^-/2}(e^{-\xn}-e^{-\x})^{1/2}]\cr
Z_2~=~&\sqrt{\pi} \theta(\x-\xnp)J_0[{1\over \sqrt{\pi}}
e^{x^-/2}(e^{-\xnp}-e^{-\x})^{1/2}]\cr}}

For $t\rightarrow -\infty$, we find  from \THREEFOUR\ that  
$(Z_2,Z_1)=-\pi$.
At an arbitrary time $t$
\eqn\THREESEVEN{\eqalign{&(Z_2,Z_1)~=~\{-\pi J_0[{1\over \sqrt{\pi}}
e^{t-\xnp/2} (e^{-\xn}-e^{-\xnp})^{1/2}]\}\cr
&-\{{\sqrt{\pi}\over 2}\int_{\xnp}^\infty dx^+ J_0[{1\over  
\sqrt{\pi}}
e^{t-\x/2}(e^{-\xnp}-e^{-\x})^{1/2}]
J_1[{1\over \sqrt{\pi}}  e^{t-\x/2}(e^{-\xn}-e^{-\x})^{1/2}]\cr
&~~~~~~~~~~~~~~~~~~~~ {e^{t-\xn}\over (e^{\x-\xn}-1)^{1/2}}\}\cr
&+\{{\sqrt{\pi}\over 2}\int_{\xnp}^\infty dx^+ J_1[{1\over  
\sqrt{\pi}}
e^{t-\x/2}(e^{-\xnp}-e^{-\x})^{1/2}]
J_0[{1\over \sqrt{\pi}}  e^{t-\x/2}(e^{-\xn}-e^{-\x})^{1/2}]\cr
&~~~~~~~~~~~~~~~~~~~~ {e^{t-\xnp}\over (e^{\x-\xnp}-1)^{1/2}}\}\cr}}
The term in the first curly brackets on the RHS (call it $C_\theta$)   
comes
from derivatives
on the theta functions; here we have used the relations $J_0(0)=1$,
$dJ_0(x)/dx=-J_1(x)$.

Let us recall the behavior of the Bessel functions $J_0$, $J_1$.  
These are
analytic functions of their argument, oscillatory for real $x$, with  
the
amplitude of oscillation decaying with increasing $|x|$. The large
$|x|$ behavior is
\eqn\THREEEIGHT{\eqalign{J_0(x)~\sim~\sqrt{{2\over \pi x}}\cos  
(x-\pi/4)\cr
J_1(x)~\sim~\sqrt{{2\over \pi x}}\sin (x-\pi/4)\cr}}

Thus $C_\theta$ vanishes for $t\rightarrow\infty$:
\eqn\THREENINE{C_\theta~\sim~ e^{-t/2}}

The other two terms in \THREESEVEN\ contain Bessel functions with
arguments containing the factor $e^t$. We find that so long as  
$\x<<2t+\xnp$
one or both Bessel functions in each term oscillates rapidly, and  
when both
factors oscillate the frequencies are equal only at isolated points.  
Using the
limits \THREEEIGHT\  and placing careful bounds on the contributions  
from
various regions of $\x$ we find that the contribution from
$\x<<2t+\xnp$ is of order $e^{-t/2}$. Thus for large $t$ we can take  
the limit
$\x\rightarrow \infty$ in $Z_1$, $Z_2$, obtaining \TWOEIGHTP\  for  
$Z_1$ and a
similar form for $Z_2$ with $\xn$ replaced by $\xnp$.  The Wronskian  
is
computed to be
\eqn\THREETEN{\eqalign{(Z_2,Z_1)~=~&-\pi\int_{-\infty}^\infty dx  
J_0[\alpha
u]J_1[u] {\partial u\over \partial t}~+~\pi\int_{-\infty}^\infty dx  
J_1[\alpha
u]J_0[u] {\partial (\alpha u)\over \partial t}\cr
=~&\pi\int_0^\infty d(\alpha u) J_0[\alpha  
u]J_1[u]~-~\pi\int_0^\infty d(\alpha
u) J_1[\alpha u]J_0[u]\cr
=~&-\pi\cr}}
as expected. Here $u={1\over \sqrt{\pi}} e^{(x^--\xn)/2}$,
$\alpha=e^{-(\xnp-\xn)/2}<1$
and we have used the relation
\eqn\THREEELEVEN{\eqalign{\int_0^\infty du J_0[a u]J_1[u]~=~&1~~{\rm
for}~a<1\cr
=~&0~~{\rm for}~a>1\cr}}

Recalling that $\xnp-\xn$ is  assumed to be not much less than unity,  
we find
that $1-\alpha$ is order unity. There is negligible contribution to  
the
integrals in \THREETEN\ for $u>>1$ or for $u<<1$ (the integrals are   
convergent
at the origin). This translates to negligible contribution to the  
Wronskian for
$|x^--\xn|>>1$, so
that the `information' is seen to come out in a width of order unity  
around
$x^-=\xn$. The latter trajectory corresponds to eikonal reflection of  
the
incident wave from the strong coupling boundary $x=0$ (with boundary  
condition
$Z=0$ at $x=0$).

Let us now see the behavior of  $Q(x)$, the `charge outside distance  
$x$'.
For the solution $Z_1$ for example, $Q(x)=0$ for $x>>t-\xn$, while
$Q(x)=\sqrt{\pi}$ for $x<<t-\xn$, with the change occuring in an  
$x$-width of
order unity around $x^-=\xn$. Thus charge comes out along the same  
trajectory
that is followed by the information in the Wronskian.

There is in addition a `blip' in the charge $Q(x)$ at the original  
trajectory
$\x=\xn$; this blip gets narrower as $t\rightarrow\infty$. If the  
initial
profile of $Z$ at  ${\cal I}^-_R$ did not have the sharp jump of the
$\theta$-function then the blip would be absent. But for a single  
fermion `in'
state we have a superposition of classical $\theta$-functions, rather  
than a
smooth classical function $Z$. Thus for such states the physical  
significance
of the blip  is not completely clear. Since the blip does not  
contribute to the
Wronskian at  $t\rightarrow\infty$ we ignore it for the purposes of  
our
discussion.

Since the Wronskian calculations are more complicated than the  
calculation of
$Q(x)$, it might be felt
that little has been gained by tracking the Wronskian to locate  
information
flow. But as later examples will show,  the charge $Q(x)$ travels  
quite
differently from The Wronskian, and the agreement of the trajectories  
in the
above case is incidental.

\subsec{$g^2=e^{2t}$.}

We saw above that spatially increasing coupling reflects  
`information' back to
the weak coupling region. But what happens to a wavepacket if the  
coupling
increases in the time direction? The theory is still unitary (though  
energy is
not conserved).  Thus the information in the wavepacket has to  
persist for
arbitrary late times, which means that it has to persist in  
arbitrarily strong
coupling regions. On the other hand we expect the field $Z$ to be  
supressed in
strong coupling regions. We wish to see how these different  
requirements are
realised by the solution $Z(t,x)$.

Thus let $g^2=e^{2t}$.  \TWOSEVEN\ is modified to
\eqn\THREETWELVE{\partial_+\partial_-Z~=~-{1\over 4\pi}g^2  
Z~=~-{1\over
4\pi}e^{x^-}e^{x^+}Z}
Let $e^{x^+}=u$, $e^{x^-}=v$. This gives
\eqn\THREETHIRTEEN{\partial_u\partial_v Z~+~{1\over 4\pi}Z~=~0}
For the initial conditions on ${\cal I}^-_R$ given by \THREEFOUR , we  
get the
solutions
\eqn\THREEFOURTEEN{\eqalign{Z_1~=~&\sqrt{\pi}  
\theta(\x-\xn)J_0[{1\over
\sqrt{\pi}} e^{x_-/2}(e^{\x}-e^{\xn})^{1/2}]\cr
Z_2~=~&\sqrt{\pi} \theta(\x-\xnp)J_0[{1\over \sqrt{\pi}}
e^{x_-/2}(e^{\x}-e^{\xnp})^{1/2}]\cr}}
with $\xnp>\xn$ as before. For $t\rightarrow -\infty$,   
$(Z_2,Z_1)=-\pi$. At an
arbitrary time $t$
\eqn\THREEFIFTEEN{\eqalign{(Z_2,Z_1)&~=~\{-\pi J_0[{1\over  
\sqrt{\pi}}
e^t(1-e^{-(\xnp-\xn)})^{1/2}]\}\cr
&-\{\sqrt{\pi} e^t\int_0^\infty dy J_0[{1\over \sqrt{\pi}}
e^t(1-e^{-y})^{1/2}]J_1[ {1\over \sqrt{\pi}} e^t(1-\alpha  
e^{-y})^{1/2}]
{1-(\alpha/2)e^{-y}\over (1-\alpha e^{-y})^{1/2}}\}\cr
&+\{\sqrt{\pi} e^t\int_0^\infty dy J_1[{1\over \sqrt{\pi}}
e^t(1-e^{-y})^{1/2}]J_0[ {1\over \sqrt{\pi}} e^t(1-\alpha  
e^{-y})^{1/2}]
{1-(1/2)e^{-y}\over (1- e^{-y})^{1/2}}\}\cr}}
We have substituted $y=\x-\xnp$, and  $\alpha=e^{-(\xnp-\xn)/2}<1$ as  
before.
For $t\rightarrow \infty$, the term in the first curly brackets goes  
to zero.

As was the case in our previous example, the remaining two terms in
\THREEFIFTEEN\ do not have significant contributions (for  
$t\rightarrow
\infty$) unless both the Bessel functions in  an integral oscillate  
slowly with
$y$. Thus there is negligible contribution for $t-y>>0$, which  
translates to
$x<<\xnp$.  Significant contributions come from a range where the  
arguments of
both Bessel functions are large, so that the expansions
\THREEEIGHT\ can be used. Then we get
\eqn\THREESIXTEEN{\eqalign{(Z_2,Z_1)~=~&-2\int_{y_0}^\infty dy~
\sin[{1\over 2 \sqrt{\pi}}e^t(1-\alpha)e^{-y}]\cr
{}~\rightarrow~&-\pi ~~~{\rm as} ~~t\rightarrow \infty\cr}}
Here $0<<y_0<<t$, with the exact choice being immaterial in the
$t\rightarrow\infty$ limit. In fact it may be seen from  
\THREESIXTEEN\ that the
essential contribution to the Wronskian
comes from  $|x-\xnp|$ order unity.  This constant $x$ trajectory  
($x=\xnp$) is
thus the location of the information at late times.

Thus we see that for a coupling varying with time as $g^2=e^{2t}$, a  
waveform
moving  left  with the speed of light at   ${\cal I}^-_R$ slows down  
to
essentially zero velocity at  $t\sim 0$ (i.e. when the coupling  
becomes $\sim
1$).  We may interpret this in terms of momentum conservation, which  
is valid
here because we have maintained translation invariance. As the  
coupling rises,
the field picks up a nonzero mass, and so the initial pulse has to  
slow down to
maintain the same momentum. We have also found that the `information'  
does not
disperse over arbitrary $x$-lengths as time progresses; rather it  
remains
confined to  a distance of order unity,
which is the scale set by the rate of rise of the coupling.  The  
values of the
functions $Z_i$ in this region tend to zero as $t\rightarrow\infty$,  
but the
frequency becomes high. In this manner the Wronskian maintains the  
value $-\pi$
even though the functions themselves `fall to zero'.

Because the information remains quite localised, we may define an  
eikonal
trajectory for the waveform  where a lightlike trajectory for $t<0$  
refracts to
a timelike (fixed $x$) trajectory at $t=0$. This behavior is  
indicated for
trajectories `C', `D', in fig. 1.

The charge $Q(x)$ is not a useful guide in the present case.  The  
total charge
on  a constant $t$ hypersurface is
\eqn\THREESEVENTEEN{Q~\equiv~\int_{-\infty}^\infty j^0(x)~=~{1\over
\sqrt{\pi}}(Z(\infty)~-~Z(-\infty))~=~ J_0[{1\over \sqrt{\pi}} e^t]}
$Q$ is unity at $t=-\infty$ but decreases to zero as  
$t\rightarrow\infty$. The
nonconservation of total charge is due to the fact that the current  
at infinity
is nonvanishing
\eqn\THREEEIGHTEEN{j^1(x=\infty)~=~{1\over \sqrt{\pi}} {\partial  
Z\over
\partial t}(x=\infty)~=~-{1\over \sqrt{\pi}}e^t J_1[{1\over  
\sqrt{\pi}} e^t]}

\subsec{Evolution in the charge trap.}

Let us now consider the coupling \ONEONE\  which forms the charge  
trap. First
we consider a trajectory like `C' in fig. 1. That is, the wavepacket  
reflects
off the first timelike segment of the strong coupling boundary, but  
then
encounters the spacelike segment of the boundary instead of escaping  
to  ${\cal
I}^+_R$. Since $g^2$  has no $\delta$-function singularities, the  
solution
$Z(t,x)$  is obtained by demanding continuity  of the solutions in  
individual
regions across the boundaries between the regions. For the theta  
function
initial conditions \THREEFOUR\ we get the complete solutions  
($\xn<0$)
\eqn\THREENINETEEN{\eqalign{Z_1~=~&\sqrt{\pi} \theta(\x-\xn)  
J_0[{1\over
\sqrt{\pi}} e^{x^-/2}(e^{-\xn}-e^{-\x})^{1/2}]~~{\rm for}~~\x<0\cr
=~&\sqrt{\pi}  J_0[{1\over \sqrt{\pi}}
e^{x^-/2}(e^{\x}+e^{-\xn}-2)^{1/2}]~~{\rm for}~~0<\x<X\cr
=~&\sqrt{\pi}  J_0[{1\over \sqrt{\pi}}
e^{x^-/2}(2e^X-2+e^{-\xn}-e^{2X}e^{-\x})^{1/2}]~~{\rm for}~~X<\x\cr}}
and $Z_2$ has the same form with $\xnp$ replacing $\xn$  
($\xn<\xnp<0$).

Let us follow for the solution $Z_1$ the charge $Q(x)$ for any fixed  
large $t$,
as $x$ is reduced from $\infty$.  $Q(x)$ vanishes for $x$ very large.
For $x\sim t+X$, $Q(x)$ decreases through zero, oscillates for a  
range of $x$
of order unity , and then settles down to $1$. Thus the fermion  
charge sent
into the charge trap gets expelled along the trajectory $x^-=-X$  
(more
precisely, in a range of width unity around this trajectory). (There  
is in
addition the expected blip in $Q(x)$, carrying zero total charge, at
$\x\sim\xn$).

Thus if we follow the charge $Q(x)$ it might appear that the  
information
emerges near $x^-=-X$. This is the trajectory of the `thunderbolt' in  
the black
hole case.  But the charge pulse we find is quite insensitive to the  
value of
$\xn$. For late times, we have $\x\rightarrow\infty$ in the  
neighbourhood of
the charge pulse, whereupon
\eqn\THREETWENTY{\eqalign{Z_1~\rightarrow~&\sqrt{\pi} J_0[{1\over  
\sqrt{\pi}}
e^{x^-/2}(2e^X-2+e^{-\xn})^{1/2}]\cr
Z_2~\rightarrow~&\sqrt{\pi} J_0[{1\over \sqrt{\pi}}
e^{x^-/2}(2e^X-2+e^{-\xnp})^{1/2}]\cr}}
{}From  fig. 1 we see that a pulse of type `C' properly enters the  
charge trap
if
\eqn\THREETWENTYP{e^{-(X+{\xn}_i)}~<<~1~~~~~~({\xn}_i~<~0)}
So long as \THREETWENTYP\ is true, the charge emerges in a width of  
order unity
around $x^-=-X$.
 So  it seems difficult for the charge pulse to carry  all the  
information of
the wavepacket sent in at  ${\cal I}^-_R$.

 We now track the contributions to the Wronskian $(Z_2,Z_1)$, and  
find that the
information actually follows quite a different path.  We find that
for late times the only region contributing to the Wronskian is where   
$x^-$
is finite, $\x=2t-x^-\rightarrow\infty$. The solutions $Z_i$ in this  
limit are
\THREETWENTY . The Wronskian calculation is the same as in \THREETEN  
, with
$u={1\over \sqrt{\pi}} e^{x^-/2}(2e^X-2+e^{-\xn})^{1/2}$ and
\eqn\THREETWENTYONE{\alpha~=~{(2e^X-2+e^{-\xnp})^{1/2}\over
(2e^X-2+e^{-\xn})^{1/2}}}

We thus recover $(Z_2,Z_1)=-\pi$. But we find that $\alpha$ is very  
close to
$1$ so  long as  \THREETWENTYP\ is true:
\eqn\THREETWENTYONEP{1-\alpha~\approx~{1\over
4}e^{-X}(e^{-\xn}-e^{-\xnp})~\sim~e^{-(X+\xn)}}

 Thus we need to be more careful in locating the essential  
contribution to the
Wronskian.  The relation \THREEELEVEN\  which we use in the Wronskian
computation is in fact discontinuous at $a=1$. To analyse this  
discontinuity
let $a$ be very close to $1$ and write the integral in \THREEELEVEN\  
as
{\eqn\THREETWENTYTWO{\eqalign{I~=~&\int_0^{u_0} du J_0[a u]J_1[u]
{}~+~\int_{u_0}^\infty du J_0[a u]J_1[u]\cr
\equiv~&I_1~+~I_2\cr}}
with $u_0=(1-a)^{-1/2}>>1$.  It is easy to see that $I_1$ gets  
contribution
only from $u$ of order unity, and we may set $a=1$ in obtaining $I_1$  
upto
corrections that vanish when $a\rightarrow 1$. In $I_2$ we use the  
asymptotic
forms \THREEEIGHT\ for $J_0$, $J_1$, to get
\eqn\THREETWENTYTHREE{\eqalign{I_2~\approx~&\int_{u_0}^\infty du  
{2\over
\pi\sqrt{a}u} \cos (au-\pi/4) \sin (u-\pi/4)\cr
\approx~&{1\over \pi}\int_{u_0}^\infty du~{\sin((1-a)u)\over  
u}~\approx~{1\over
2}~{\rm sign}~(1-a)\cr}}
{}From \THREEELEVEN\ we see that $I_1=1/2$.

Using the breakup \THREETWENTYTWO\ in the Wronskian calculation we  
find that
the contributions of type $I_1$ cancel while those of type $I_2$ add  
to give
$-\pi~ {\rm sign}~(1-\alpha)$. The contributions to $I_2$ in  
\THREETWENTYTWO\
come from $u\sim (1-a)^{-1}$, which translates for the Wronskian  
calculation to
\eqn\THREETWENTYTWOP{x^-~\approx~X~+~2\xn}
 This trajectory is quite different from the trajectory of the charge  
found
above, which was $x^-=-X$. Further, the contributions to the $I_2$  
integral in
\THREETWENTYTWO\ come from a {\it factor} of order unity around
 $u\sim (1-a)^{-1}$, which translates to an $x$-width of order unity  
around the
trajectory  \THREETWENTYTWOP\ for the contribution to the Wronskian.

To summarise, the information in a wavepacket of type `C' in fig. 1  
will  first
reflect from the timelike line $x=0$. It will  then enter the region  
where the
coupling grows with $t$ where it travels at essentially constant $x$  
after
$t=0$. It  then encounters spatially varying coupling again and  
emerges to
${\cal I}^+_R$. The emerging trajectory is retarded by time $X+\xn$  
compared to
the case where the strong coupling boundary was $x=0$ for all $t$  
(i.e. no
charge trap).  By contrast, the charge of the wavepacket is expelled  
to ${\cal
I}^-_R$ along a track advanced by $X+\xn$ compared to the `no charge  
trap'
case. The eikonal approximations to the above trajectories are  
sketched in fig.
1.

In a similar manner we can analyse the wavepacket marked `D' in fig.  
1, where
the  pulse first hits the spacelike section of the strong coupling  
boundary.
The solutions with boundary condition \THREEFOUR\ are ($0<\xn<X$)
\eqn\THREETWENTYFOUR{\eqalign{Z_1~=~&\sqrt{\pi}\theta(\x-\xn)J_0[{1\o 
ver
\sqrt{\pi}} e^{x^-/2}(e^{\x}-e^{\xn})^{1/2}] ~~{\rm for}~\x<X\cr
=~&\sqrt{\pi}J_0[{1\over \sqrt{\pi}}\sqrt{\pi}
e^{x^-/2}(2e^X-e^{\xn}-e^{2X}e^{-\x})^{1/2}] ~~{\rm for}~X<\x\cr}}
and $Z_2$ is defined similarily with $\xnp$ replacing $\xn$  
($0<\xn<\xnp<X$).

Note that if we let $X\rightarrow 0$ we do not recover  \THREESIX ,  
the
solution with the purely timelike boundary. But this is not a  
contradiction,
because the solution \THREETWENTYFOUR\ is arranged to have the charge  
hit the
spacelike part of the boundary, however small that may be. The
solution\THREENINETEEN\ does go over to the form in \THREESIX\ when
$X\rightarrow 0$.

As in the above case, the charge is expelled along $x^-=-X$.  The   
Wronskian
gets its contribution from a unit $x$-width around
\eqn\THREETWENTYTWOPP{x^-~\approx~X~-~2\xn}
In this calculation the condition     \THREETWENTYP\  is replaced by
\eqn\THREETWENTYPP{e^{-(X-{\xn}_i)}~<<~1~~~~~~(0~<~{\xn}_i~<~X)}
This condition is again just the requirement that the pulse enter the  
charge
trap.

We note that in all the cases studied here the location of the  
contribution to
the Wronskian is at the eikonal trajectory which is at the larger $x$  
value at
the time $t$ under consideration, just as in the computation  
\THREEFOURP .

Following the calculation in \PST\ we can compute the $S$ matrix
for the charge trap
\eqn\FOURFOUR{\eqalign{S~=&~{\rm exp}\{i\pi\int dx^+dx^-j_R(x^-)\{
\theta(-\x)J_0[{1\over \sqrt{\pi}}  
e^{x^-/2}(2e^X~-~2~+~e^{-\x})^{1/2}]\cr
&~~~~+ (\theta(\x)-\theta(\x-X))J_0[{1\over \sqrt{\pi}}
e^{x^-/2}(2e^X~-~e^{\x})^{1/2}]\cr
&~~~~+\theta(\x-X)J_0[{1\over \sqrt{\pi}} e^{x^-/2}e^{X-\x/2}]\}
j_L(\x)\}}}

\newsec{Discussion.}

Our goal was to study a very simple model where an incident  
wavepacket meets a
spacelike boundary of strong coupling, so that the `information' is  
forced to
enter the strong coupling region. This would parallel the situation  
encountered
in the black hole. Let us compare our results with that for the  
charge hole,
where the strong coupling boundary is timelike. In the latter case  
the
information   of the incident charge pulse is carried by   neutral
fermion-antifermion pairs that follow the escaping charge to
 ${\cal I}^+_R$. The information coded into  high frequency modes of  
the
incident pulse comes out  further down the tail of the outgoing wave,  
but the
information in wavelengths order unity travels within unit distance  
of the
charge carried by the outgoing wave.

 When wavepackets of type `C' or `D' encounter the spacelike boundary  
of the
`charge trap' the situation is somewhat different. Even when the  
frequency of
the incident wave is not high, the charge and the information  
carrying neutral
pairs get separated. The  amount of separation depends on the size  
$X$ of the
charge trap and the position of the incident wave with respect to the  
trap, but
is practically independent of the wavelength of the incident wave so  
long as
this is not much smaller than unity or approaching the size $X$ of  
the trap. An
interesting feature is that the `information' as described by the  
Wronskian
remains well localised as it moves thriough the strong coupling  
region and out
to  ${\cal I}^+_R$;  it does not spread to $x$-widths larger than  
order unity.
This feature allows us to draw the `eikonal trajectories' for  
information flow
given in fig. 1. Thus we see that the location of information at   
${\cal
I}^+_R$ is not monotonic in its initial position on  ${\cal I}^-_R$.
 (We find  that these locations are related by $x^-=X-2|\x|$.) The  
`refraction'
 of the wave through the strong coupling region cannot be replaced by  
a simple
reflection off any boundary.

The charge in the incident pulse is expelled at $x^-=-X$. We can  
interpret this
as follows. If the incident pulse has positive charge, then we should  
imagine
an equal negative charge remaining at $x\rightarrow\infty$ so that  
there is no
net electric field for $x\rightarrow-\infty$. The latter condition is  
required
to avoid having infinite electric field energy in the strong coupling  
region.
But there is nonvanishing field to the right of the charge pulse, so  
that when
the coupling becomes strong at the spacelike line $t=0$ ($x<X$)   
fermion
anti-fermion pairs are quickly produced to neutralise the field. The  
charge is
thus pushed to $x=X$   `acausally'.
(Of course this is not a contradiction because  charge is a globally  
conserved
quantity.) The information, on the other hand,  must move within the
characteristics of the wave equation, so it is satisfactory that it  
comes out
retarded instead of advanced in time.

It may be possible to construct models with $g^2(t,x)$ such that the
information never  escapes to ${\cal I}^+_R$. (It could for example  
move to
other asymptotic regions as argued for the Reissner-Nordstrom black  
hole \TRI .
)  But let us note here the basic reason why the information had to  
come out in
our simple model.  For $t\rightarrow\infty$ we want to have, to an  
arbitrarily
good approximation, the time independent vacuum with
$g^2=e^{-2(x-X)}$. Thus for late times this form of the coupling must  
extend to
$x<<0$. But the incident pulse, on  reaching the spacelike part of  
the strong
coupling boundary, travelled at essentially constant $x$.  Thus at  
some time in
the future this pulse must encounter a region where $g^2$ is  
decreasing with
$x$, whereupon it  would escape to   ${\cal I}^+_R$.

The analogue of our result for the gravitaional black hole would be  
that the
Hawking radiation carries off all the mass, but the information leaks  
out later
from the temporary remnant. Thus the information gets buried deep in  
the strong
coupling region, perhaps forming a   `cornucopion' \BD , but when the  
hole
loses its mass by evaporation then the `cornucopion' expels the  
information out
instead of keeping it forever in a remnant. If this picture were  
correct for
the black hole then we should not be looking for the information in  
the Hawking
radiation.

There are of course many features of the black hole problem that are  
not
reflected in the charge model. We have no fluctuations of the metric  
which
could violate naive causality. The dilaton is also taken as classical  
and
fixed, rather than as a quantum field. The structure of the charge  
trap is not
self-consistently determined by the infalling matter, but is fixed to  
have its
strong coupling boundary resemble that of a black hole.  Finally, the
gravitational coupling is of course quite different from the  
electromagnetic
one. For instance we expect gravity to treat all fields the same way,  
while in
the charge hole if we have many species of fermions, then only one  
combination
of currents if reflected back to the weak coulping region \PST .

In spite of these differences it appears concievable that one could  
make a
model of gravity (perhaps by adding higher derivative terms) such  
that the
strong coupling region  has no singularities in the metric or  
dilaton.    We
can consider the path of information carried by a test particle  
thrown in after
the hole forms, such that this particle has a small influence on the  
geometry.
If the strong coupling domain indeed reflects information  then we  
might  find
the situation that we encountered in the charge trap, which is that  
information
emerges over time  from the temporary remnant.

One objection to information recovery from a small remnant is the  
following
\ref\AHAR{Y. Aharonov, A. Casher and S. Nussinov, Phys.  Lett. {\bf B  
191}
(1987) 51.}.  For the black hole, the charge in the charge hole must  
be
replaced by mass, in all arguments.  Thus the remnant holding the  
information
will have low mass. The remnant must decay into  a large number of  
quanta to
carry out all the information. But since the total mass available is  
small, the
quanta have very long wavelengths, which means that the overlap  
amplitude of
the remnant with its decay products is very small. Thus very long  
times are
needed to encode information into the outgoing particles, so that the
information is effectively stored forever. (If the black hole mass is  
more than
a few planck masses, the decay time is estimated to exceed the age of  
the
Universe.)

But such a heuristic argument may not be accurate for the end state  
of black
hole evaporation.
 Unlike the classical stress tensor for, say, a massless scalar  
field, the
quantum stress tensor need not have positive expectation value of  
energy
density everywhere. While the total energy must be positive, regions  
of
negative energy density are produced if we consider squeezed states,  
instead of
coherent states. Squeezed states have the form $\sim e^{\mu a^\dagger
a^\dagger}|0>$, and are pertinent to pair creation processes.
 Thus it might be possible to have the mass come out in the Hawking  
radiation
and the information to come out  later in  a region with no  
expectation value
of the stress tensor.

 It has been suggested however that even though the mean energy can  
be made to
vanish in a region in this fashion, the price to be paid is that  
there are
large fluctuations in the energy about the mean \ref\FORD{L.H. Ford  
and
C-I Kuo, Preprint TUTP-93-1 (gr-qc 9304008) (1993)}\ref\TRIPRES{S.  
Trivedi and
J. Preskill, Private communucation by S. Trivedi.}. The root mean  
square
fluctuation of the energy in such a late time `information carrying'  
part of
the state would be of the order of the black hole mass. Is it  
physically
reasonable to expect such large fluctuations when we know that the
semiclassical calculation  tells us that practically all the mass  
comes out in
the Hawking radiation?

 This appears to be a complicated issue. The semiclassical analysis  
may
 be trustworthy for expectation values of the stress energy, which  
should
therefore vanish after the Hawking radiation comes out. To estimate  
the
fluctuations in stress energy may need a complete quantum gravity  
treatment,
because
 the part of the state passing through the strong coupling region  
would be
  influenced by    loops of  all order in the matter - gravity  
interaction.
How much fluctuation in stress energy can the resulting state have?  
In a simple
quantum mechanical system, the expectation value of the energy and  
its
dispersion are both conserved as the state evolves, because all  
moments of the
Hamiltonian are preserved.  But in quantum gravity,  we have that the  
total
Hamiltonian is zero on the state.\foot{Difficulties asscociated with  
this have
been discussed in the 2-dimensional context in \ref\DEAL{S. P.  
deAlwis,
Preprint COLO-HEP-318 (hep-th 9307140).}} The dispersion in this  
total
Hamiltonian thus vanishes, but there is no reason for the dispersion  
in the
matter stress energy to be separately conserved. In fact to make  
sense of the
semiclassical limit of  space-time one needs to use the notion of
`decoherence', which may break down in the strong coupling region of  
the black
hole.  It is concievable that matter states with large fluctuations  
in stress
energy are created in such a process; we hope to study this aspect  
further.

\bigskip
\bigskip
\centerline{\bf Acknowledgements}
\bigskip
I wish to thank for useful discussions S. Axelrod,  R. Brooks, J.  
Cohn, M.
Crescimanno,  A. Dabholkar, D. Freed   D. Kabat, J. Preskill and S.  
Trivedi.
This work is supported in part by DOE grant DE-AC02-76ER.

\bigskip
\listrefs
\bigskip
\bigskip
{\bf Figure Caption:}
Fig. 1: \quad  The trajectories of `information flow' in the `charge  
trap'. The
heavy line is the strong coupling boundary $g^2=1$.
The charge escapes along  the line marked with an arrow.

\bye